\documentstyle[aps,multicol,epsfig,prb,floats]{revtex}
\newcommand{\bef}{\begin{figure}}
\newcommand{\enf}{\end{figure}}
\newcommand{\bec}{\begin{center}}
\newcommand{\enc}{\end{center}}

\newcommand{\Vg}{V_{\mbox{\scriptsize g}}}

\begin{document}
\draft
\bibliographystyle{prsty}
\twocolumn[\hsize\textwidth\columnwidth\hsize\csname @twocolumnfalse\endcsname
\title{Temperature dependence of Fano line shapes in a weakly coupled single-electron transistor}
\author{I. G. Zacharia, D. Goldhaber-Gordon\cite{davidaddress}, G. Granger, M. A. Kastner\cite{byline}, and Yu. B. Khavin}
\address{
Department of Physics, Massachusetts Institute of Technology\\
Cambridge, MA 02139
}
\author{Hadas Shtrikman, D. Mahalu, and U. Meirav}
\address{
Braun Center for Submicron Research\\
Weizmann Institute of Science\\
Rehovot, Israel 76100
}

\maketitle

\begin{abstract}
We report the temperature dependence of the zero-bias conductance of a sin\-gle-electron transistor in the regime of weak coupling between the quantum dot and the leads.  The Fano line shape, convoluted with thermal broadening, provides a good fit to the observed asymmetric Coulomb charging peaks.  However, the width of the peaks increases more rapidly than expected from the thermal broadening of the Fermi distribution in a temperature range for which Fano interference is unaffected. The intrinsic width of the resonance extracted from the fits increases approximately quadratically with temperature.  Above about 600~mK the asymmetry of the peaks decreases, suggesting that phase coherence necessary for Fano inter\-ference is reduced.
\end{abstract}


]

A single-electron transistor (SET) consists of a small, isolated conductor, coupled to metallic leads by tunnel junctions. The confinement quantizes the charge and energy of the isolated region, making it closely analogous to an atom~\cite{kastner,ashoori}. For such structures the conductance, resulting from transmission of electrons from one lead to the other, consists of peaks as a function of gate voltage, one for each electron added to the artificial atom.  The peaks occur when two charge states of the artificial atom are degenerate in energy, at which point resonant tunneling can occur at zero temperature.  Between the peaks the conductance at low temperature is expected to be limited by virtual excitations of electrons on and off the artificial atom, a non-resonant process called co-tunneling~\cite{averin}.

G\protect{\"{o}}res {\it et al.}~\cite{Gores} have recently reported Fano line shapes in the conductance peaks for a small SET.  This implies that there are two paths through the SET at each energy, one resonant and the other non-resonant, that interfere with each other.  G\protect{\"{o}}res {\it et al.} have exam\-ined the Fano interference for the case when the coupling to the leads is strong, and the non-resonant contribution to the conductance is then comparable in size to the resonant component. We here report the observation of Fano line shapes when the coupling is weak and the non-resonant conductance is small.  We find that the Fano functional form, broad\-ened by the Fermi-Dirac distribution function, provides a good fit to the line shape between $100$ and $800$~mK.  With increasing temperature $T$ the intrinsic width of the resonance increases, approximately quadratically with $T$. This increase is reminiscent of that expected from inelastic scattering, but it is more rapid and occurs at temperatures for which Fano interference is apparently unaffected. Above $\sim600$ mK the asymmetry of the peaks decreases more rapidly than predicted from thermal broadening alone, suggesting that phase coherence is destroyed with increasing T.

\begin{figure}[hbt]
\epsfig{file=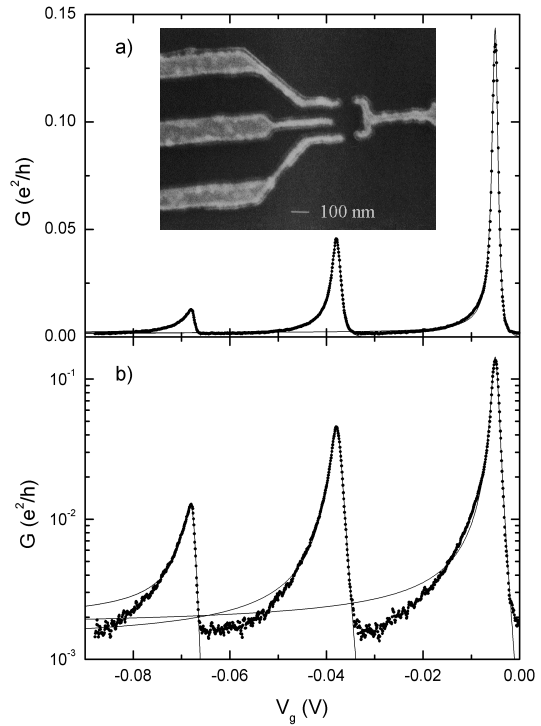, height=0.62\textwidth, width=0.48\textwidth}
\caption{Conductance as a function of gate voltage (a) on a linear conductance scale and (b) on a logarithmic scale.  The solid curves are 4 param fits as described in the text using Eqs.~\ref{fermi} and~\ref{fano} and $T=80$ mK.  The parameters found from the fits for the three peaks, from right to left, respectively, are $q= -9.33$, $-5.87$, and $-2.41$; $\Gamma=0.225$~meV, $0.375$~meV, and $0.372$~meV; $G_0=1.65\times 10^{-3}$~$e^2/h$, $1.27\times 10^{-3}$~$e^2/h$, and $1.87\times 10^{-3}$~$e^2/h$.}
\label{figure1}
\end{figure}

The SETs we have studied are similar to the ones used by Goldhaber-Gordon
$\it{et\ al.}$ to study the Kondo effect~\cite{david,davidprl}. The SET is
created by imposing an external potential on a two-dimensional electron gas (2DEG) at
the interface of a GaAs/AlAs heterostructure.  Our 2DEG has a mobility of
$91,000\ \mbox{cm}^2/\mbox{Vs} $ and a density of
$7.3\times 10^{11}\ \mbox{cm}^{-2}$; these quantities have been measured shortly after fabrication and may change somewhat with time. We create the confining potential with electrodes shown in the inset of Fig.~\ref{figure1}, in combination with a shallow etch beneath the gates. Applying a negative
voltage to two pairs of split
electrodes depletes the 2DEG underneath them and
forms two tunnel barriers separating a droplet of electrons from the 2DEG regions on either side, which act as the source and drain leads. The confinement caused by the electrodes is supplemented by shallow etching before the gate electrodes are deposited.  Our SETs are made with a 2DEG that is closer to the surface
($\approx 20$ nm) than in most other experiments, allowing the electron
droplet to be confined to smaller dimensions. This also makes the tunnel
barriers more abrupt than in previous structures. We estimate that
our droplet is about $100$ nm in diameter and contains about $50$ electrons.

The electrochemical potential of the electrons on the droplet can
be shifted relative to the Fermi energies
in the leads using an additional plunger gate electrode near the droplet.  Throughout this paper we discuss measurements in which the split electrode voltages are fixed, and we refer to the voltage on the ``plunger" as the gate voltage $\Vg$.  We measure the conductance by applying a small alternating 
voltage (typically $7\mu$V) between the drain and source leads and measuring the current with a current preamplifier and a lock-in amplifier. With such excitation voltages, less than $kT/e$, we are confident that we are measuring the zero-bias conductance of the SET.  The conductance is then recorded
as a function of $\Vg$.

Depending on the transmission of the tunnel barriers, we observe
different transport regimes in our SETs.
As is often observed in semiconductor SETs, each time we cool a particular SET to low temperatures, we find different tunneling rates of the barriers and a different electrochemical potential of the electron droplet, for the same electrode voltages.
This probably reflects the metastability, at low temperature, of electrons in the donor layer within the AlAs. Thus, the same SET that shows strong coupling in one cool down may show weak coupling in another.  Also possibly related to this metastability are events in which the effective voltage of a particular electrode suddenly changes to a different value.  We suspect that this switching behavior results from charge motion around an impurity or defect near the artificial atom.  The switching can change the SET's characteristics and can therefore limit the degree to which we can study them before they change.

Figure~\ref{figure1} shows the conductance as a function of $\Vg$ for a situation in which the coupling to the leads is weak, as evinced by the the small height of the single-electron peaks.  The peak widths are also relatively small, as discussed below.  It is evident from the plot on a linear scale in Fig.~\ref{figure1}a that the peaks are asymmetric, but this is seen even more clearly in the logarithmic plot of Fig.~\ref{figure1}b.  We have observed such asymmetric line shapes in different coupling regimes for about 25 peaks in five SETs.

The Fano formula~\cite{fano} gives asymmetric resonances similar to those in Fig.~\ref{figure1}; it describes the transmission resulting from interference of a resonant channel at energy $\epsilon_0$ with a non-resonant channel.  While G\protect{\"{o}}res {\it et al.}~\cite{Gores} used the zero-temperature Fano cross-section to model their results, we here include thermal effects.  For source-drain voltages $\ll kT$ the conductance is related to the transmission coefficient $\Theta(\epsilon)$ at energy $\epsilon$ by
\begin{equation}
G= \frac{e^2}{h} \int d\epsilon \Theta(\epsilon)
\frac{\cosh ^{-2} (\frac{\epsilon-\mu}{2kT})}{4kT}
\label{fermi}
\end{equation}
where  $\mu$ is the chemical potential.  Introducing the Fano form for the transmission, we write
\begin{equation}
\Theta(\epsilon)=\Theta_0
\frac{
(\tilde{\epsilon}+q)^2}
{\tilde{\epsilon}^2 + 1}
\label{fano}
\end{equation}
where the dimensionless detuning from resonance is $\tilde{\epsilon} \equiv (\epsilon - \epsilon_0)/(\Gamma/2)$. $\Gamma$ is the width of the resonance and $\Theta_0$ is the transmission far away from the resonance.
The parameter $q$ is proportional to the ratio of transmission amplitudes for the resonant and non-resonant channels~\cite{fano}; the sign of q depends on the phase shift between the two channels.   For $|q| \rightarrow 0$ non-resonant transmission dominates, resulting in a symmetric dip at the resonant position. For $q\simeq1$ the peak is highly asymmetric.  If there is no non-resonant transmission, $|q| \rightarrow \infty$ and $\Theta_0 \rightarrow 0$ in such a way that $\Theta_0|q|^2$ approaches a constant leading to a Breit-Wigner line shape.  

One can relate $|q|$ to the parameters characterizing the quantum dot by considering the Breit-Wigner limit at $T=0$. One demands that the relationship of $\Theta_0q^2$ to $\Gamma_R$ and $\Gamma_L$, the energies resulting from tunneling of electrons from the artifi\-cial atom to the right and left leads, respectively, be in agreement with the usual result for peak heights in SETs with no continuous transmission channel. When $kT\ll\Gamma\ll\Delta \epsilon$, where $\Gamma=\Gamma_R+\Gamma_L$ and $\Delta \epsilon$ is the level spacing, this gives~\cite{beenakker}
\begin{equation}
|q|\simeq\frac{1}{\Gamma_R+\Gamma_L}2\sqrt{\Gamma_R \Gamma_L/\Theta_0}
\label{sven}
\end{equation}
The Breit-Wigner peak then has maximum conductance, in units of $e^2/h$, of $\Theta_0 q^2 \sim 4 \Gamma_R\Gamma_L/(\Gamma_R+\Gamma_L)^2$, the value expected for single-level transport through an SET.~\cite{beenakker}  Conventional single-electron charging peaks are characterized by three parameters--the peak position and the couplings to the two leads.  A fourth parameter is required to include the amplitude of the non-resonant channel.

Changes in gate voltage are proportional to changes in chemical potential $\mu$; specifically, $\mu = \alpha e\Vg$, apart from an additive constant.  The parameter $\alpha$ may be determined from a measurement of the various capacitances between the artificial atom and its nearby leads and gates:  $\alpha = C_g/C_{tot}$, where $C_g$ is the capacitance between the quantum dot and the plunger gate, and $C_{tot}$ is the total capacitance. From previous measurements of $C_g/C_{tot}$ on our SETs in the well-isolated regime, we find that $\alpha\simeq 0.15$.  However, for some of our fits we have measured $\epsilon$ and $\Gamma$ in units of gate voltage, leaving $\alpha$ undetermined.

Using Eqs.~\ref{fermi} and~\ref{fano} we have fit the data in three ways in order to demonstrate that the temperature dependences of the important parameters do not depend on the fitting procedure.  First, we set $T=0$ in Eq.~\ref{fermi}, so that the Fano form is used at all $T$, and take $q$, $\Gamma$, $G_0=\Theta_0e^2/h$, and $\epsilon_0$ as parameters. We call this procedure, followed by G\protect{\"{o}}res {\it et al.}, ``Fano''.  It obviously leads to a temperature dependent $\Gamma$ because it includes no broadening of the electron distribution.  Second, we also allow $kT/\alpha e$ to vary,  giving us five parameters instead of four, and we refer to this as ``5 param''.  We do this to check that our value of $\alpha$ is consistent with the observations.  Third, we fix $\alpha = 0.15$, the value found by Goldhaber-Gordon {\it et al.}~in the well-isolated regime, and, using the measured value of $T$, we  allow $q$, $\Gamma$, $G_0$, and $\epsilon_0$ to vary; we call this ``4 param''.  For all three approaches we have fit the logarithm of the conductance, rather than the conductance itself, to give appropriate weight to the asymmetric tails. We fit over the range for which the conductance is at least twice as large as the lowest conductance measured, in order to minimize the effect of neighboring peaks.  The agreement of the parameters determined by the three fits gives us confidence in the main conclusions we draw.

We show the 4 param fit for the three peaks in Fig.~\ref{figure1}.  We have fit the three separately, because we know of no way to fit multiple Fano resonances whose tails overlap.  G\protect{\"{o}}res {\it et al.}~\cite{Gores} have found an incoherent contribution to the non-resonant conduc\-tance, comparable to the coherent part at all temperatures. We have tried adding such an incoherent component, and find that it has about half the magnitude of the coherent part, but is negative.  Because of this unphysical result and because including the incoherent component does not improve the fits appreciably, we have set it to zero for all fits shown here.  

Note that one of the values of $|q|$ in Fig.~\ref{figure1} is as large as $\sim10$, giving a ratio of peak-to-background conductances of $\sim100$, whereas G\protect{\"{o}}res {\it et al.}~found $|q|$ of order one, consistent with their observation that the coherent non-resonant component is of the same size as the resonant one.

The sign of $q$ reflects the average phase shift between the continuous transmission channel and the resonant channel.  This will certainly be random from one sample to another and even from one peak to another in the same sample.
Although $q$ is negative for the peaks in Fig.~\ref{figure1}, we have found other peaks for which $q$ is positive.  Indeed, Goldhaber-Gordon {\it et al.}~have reported data, also in the weakly coupled regime like those discussed here, that have the opposite asymmetry from those shown here~\cite{david}.

\begin{figure}[hbt]
\epsfig{file=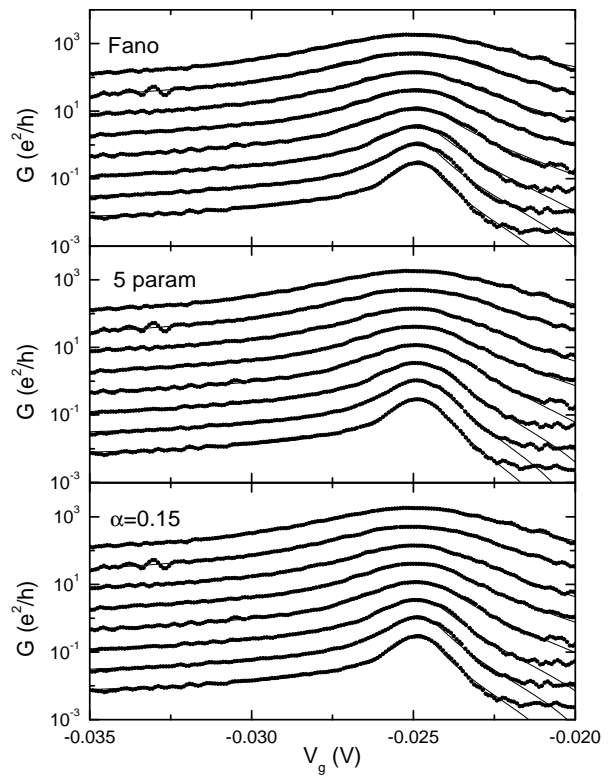, height=0.62\textwidth, width=0.48\textwidth}
\caption{Data for one conductance peak taken in $100$ mK steps between $100$mK (lowest) and $800$mK (highest).  The same data are fit in each of the three panels in different ways as discussed in the text.  The data are multiplied by a factor $\sqrt{10}$ from one set of data to the next.}
\label{figure2}
\end{figure}

Figure~\ref{figure2} shows the evolution of a single conductance peak with temperature together with the results of the three fits on logarithmic scales.  The resonance position $\epsilon_0$ varies randomly by a small amount from one temperature to another, presumably because of the switching phenomena discussed above; for all fits we have shifted the curves so that their resonance energies coincide with that for the $100$~mK data.

\begin{figure}[hbt]
\epsfig{file=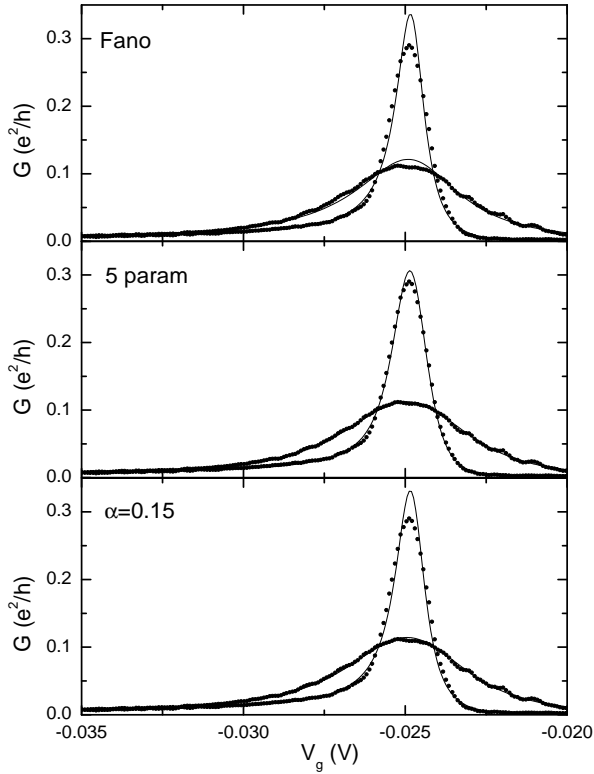, height=0.62\textwidth, width=0.48\textwidth}
\caption{Examples of the data and fits from Figure~\ref{figure2} on linear conductance scales.  Each panel contains results for $100$ mK and $800$ mK.}
\label{figure3}
\end{figure}

\begin{figure}[hbt]
\epsfig{file=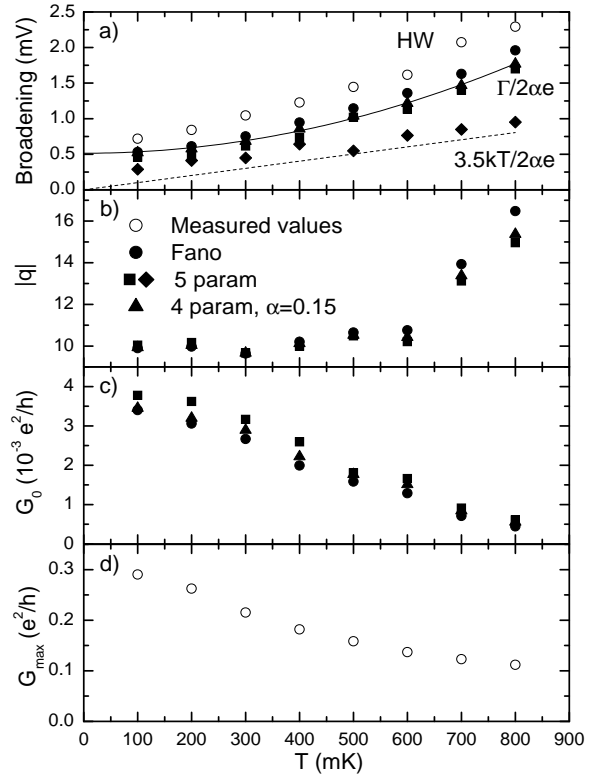, height=0.62\textwidth, width=0.48\textwidth}
\caption{Parameters extracted from the fits in Fig.~\ref{figure2}.  The open circles are the measured half width (HW) in (a) and peak height in (d).  The solid circles, triangles, and squares are the parameters from the fits using Fano, 4 param, and 5 param, respectively.  The diamonds are the temperature broadening, $3.5kT/2\alpha e$, extracted from the 5 param fit, and the dashed line is the same broadening expected for $\alpha=0.15$.  Note that the values of $\Gamma$, $|q|$ and $G_0$ are independent of the fitting method used.}
\label{figure4}
\end{figure}

It is clear that all three fits are excellent except near the conductance minimum, on the right side of the peak.  As seen in Fig.~\ref{figure3}, a linear plot of data selected from Fig.~\ref{figure2}, the fits overestimate the peak height slightly at low $T$.  This discrepancy may be minimized by fitting to the conductance instead of its logarithm, but then the fit in the  tails is poor.  Overall, the 5 param fit is slightly better than the other two, but the values of the parameters are essentially the same, as seen in Figure~\ref{figure4}.

This figure shows the measured half width (HW) and the conductance maximum $G_{max}$.  The HW increases from about $0.5$~mV at $100$~mK to nearly $2.5$~mV at $800$~mK.  The peak height decreases by about a factor $2$ between $100$ and $600$ mK, and then more slowly at higher $T$.  Equation~\ref{fermi} predicts that the HW increases linearly with $T$, with a finite zero-temperature intercept, related to $\Gamma$.  The increase of width predicted for the latter case is shown by the dashed line in Fig.~\ref{figure4}, HW$=3.5kT/2\alpha e$.  Equation~\ref{fermi} also predicts that, at $T$ high enough that the HW is much larger than its $T=0$ value, the peak height decreases as $T^{-1}$.  Thus, the measured HW increases more rapidly than expected, and the peak height decreases more slowly.

Turning to the fitting parameters, we first notice that $|q|$ is approximately constant up to $\sim600$ mK and then increases rapidly.  The increase of $|q|$ reflects an increasing symmetry of the line shape, even though the HW is still larger than $3.5kT/2\alpha e$.  Furthermore, $G_0$ becomes very small at these high temperatures, while $G_{max}\simeq G_0q^2$ approaches a constant.  Thus, the line shape approaches a Lorentzian at the highest temperatures, suggesting a loss of the phase coherence required for Fano interference.   

Second, we observe that $\Gamma$ increases with $T$ even in a temperature range for which $|q|$ is constant.  Such a broadening of the intrinsic width of Coulomb charging peaks has been predicted to arise from inelastic scattering~\cite{stone,buttiker}. In applying this to a SET, Beenakker~\cite{beenakker} has proposed that the Breit-Wigner form be written
\begin{equation}
G_{BW}=g\frac{e^2}{h}\frac{\Gamma_R\Gamma_L}{\Gamma_R+\Gamma_L}\frac{\Gamma}{\epsilon^2+(\Gamma/2)^2}
\label{been}
\end{equation}
where $\Gamma$ now contains both elastic and inelastic components.  Thus, an increase of $\Gamma$ with temperature is not unexpected.  However, we find it surprising that the lifetime of electrons in the quantum dot decreases considerably with increasing temperature, at temperatures for which there is still phase coherence between the continuous and resonant tunneling channels. 

We find that $\Gamma$ increases approximately quadratically with $T$; the fit shown in Fig.~\ref{figure4}a yields $\Gamma=((0.15\pm0.01)+(0.59\pm0.01)T^2)$~meV, with $T$ in kelvins.  According to Sivan {\it et al.}\cite{sivan}, since our mean free path (of order one micron) is much larger than the dimensions of our quantum dot, and the energy scale of our measurement ($kT$) is much smaller than $\Delta\epsilon$, our system is in the clean limit for which $\Gamma= \Gamma_0+(\pi^2/8)(kT)^2/E_F\sqrt{\pi k_F a_B}$, where $\Gamma_0$ is the zero temperature value, $k_F$ is the Fermi wavevector in the 2DEG, and $a_B$ is the effective Bohr radius.  This predicts that the increase of $\Gamma$ with $T$ should be quadratic, but the increase should be much less in magnitude than $kT$, as usual for a Fermi liquid.  Consistent with this, Huibers {\it et al.}~\cite {huibers} find phase coherence times for open quantum dots that correspond to broadening that is less than $kT$ for the temperature range studied.  Our observation of $\Gamma$ values that are considerably larger than $kT$ is thus very surprising.  

For the peak studied in Fig.~\ref{figure2}, the contribution to the broadening at $\sim600$ mK from the Fermi-Dirac distribution is about the same size as that from the increase in $\Gamma$.  Thus, examination of the measured HW of the peak in Fig.~\ref{figure4}a shows that it is difficult, without the fitting procedure used here, to distinguish the temperature dependence below $600$ mK from the usual Fermi-Dirac broadening.  Our group has reported~\cite{Gores} that the broadening of Fano peaks in the strongly coupled regime is approximately that predicted by Fermi-Dirac up to $\sim1$ K, but is larger than expected at higher $T$.  In light of the present measurements, we suggest that the HW reported by G\protect{\"{o}}res {\it et al.} may also be influenced by a strongly temperature dependent $\Gamma$.

We have shown here the temperature evolution of one of our largest peaks. We have also studied a smaller one.  Like the smaller peaks in Fig.~\ref{figure1}, the smaller peak has a larger width and a smaller value of $|q|$.  For the smaller peak $q$ is more sensitive to $T$ at low $T$.  Presumably, when $|q|$ is closer to unity, it is more sensitive to a decrease in phase coherence. We find that the width of this peak is less temperature dependent than that of the larger peak. This may be a consequence of its width being larger even in the limit
$T\rightarrow0$.

G\protect{\"{o}}res {\it et al.}~\cite{Gores} have measured the source-drain voltage dependence as well as the gate voltage dependence of the peaks in differential conductance.  The results show the diamond structure characteristic of single-electron charging.  Thus, the nature of the resonant contri\-bution to the conductance is clearly the same as in other SETs.  However, the origin of the continuous channel is more difficult to identify.  Because G\protect{\"{o}}res {\it et al.} observe a continuous contribution that is of order $e^2/h$, they have suggested that the continuous channel results from transmission at energies higher than the potential barriers.  However, for the regime examined here, in which the background conductance is $\sim10^{-3}$~$e^2/h$, that does not seem reasonable.

Another possibility raised by G\protect{\"{o}}res {\it et al.} is that the coherent non-resonant component arises from co-tunneling.  Co-tunneling results in a continuous contribution to the conductance in the valley between peaks that is stongly dependent on gate voltage.~\cite{averin}  Thus, it is surprising that $G_0$ is independent of gate voltage.  Furthermore, since the co-tunneling rate depends on $\Gamma_L$ and $\Gamma_R$, which vary from peak to peak, one would expect a contribution to the continuous component near a given peak to vary with the heights of neighboring peaks.  Unlike the situation studied by G\protect{\"{o}}res {\it et al.}, our peaks vary in magnitude by a factor $\sim10$.  Therefore, one would have expected a large variation in the con\-tinuous contribution from one peak to the next.  However, returning to Fig.~\ref{figure1} we note that the continuous contribution to the conductance for all three peaks is the same within the errors, $G_0=(0.0016\pm0.0003)$~$e^2/h$.  We conclude that co-tunneling is not a likely source of the continuous background.

The resonant component we observe appears to be unrelated to that giving rise to Fano line shapes observed by Li {\it et al.}~\cite{li} and Madhavan {\it et al.}~\cite{madhavan}.  The latter authors ascribe the resonant component to the Kondo singlet formed when a localized spin is screened by the spins of electrons in nearby metallic leads.  From the detailed gate voltage and temperature dependencies of the Kondo effect reported by Goldhaber-Gordon {\it et al.}~\cite{davidprl}, we know that the Kondo effect results in a {\it decrease} in peak width with increasing temperature, the opposite of what we observe.  Furthermore, Goldhaber-Gordon {\it et al.} observe a clear alternation in peak behavior, depending on whether the peak corresponds to a change from even to odd or from odd to even occupancy of the electron droplet.  From Fig.~1, it is obvious that no such effects are seen in the present experiments.  We believe that the tunnel barriers for the situation studied here are large so that the Kondo temperature is much smaller than our base temperature for all gate voltages.  Thus the resonances we observe are probably the bare ones associated with the energy levels of the artificial atom.

In summary, we have found that the Coulomb charging peaks in small SETs are asym\-metric and are well described by the Fano line shape even when the coupling to the leads is weak.  It is not surprising that increasing the temperature destroys the phase coherence required for the Fano interference.  However, it {\it is} surprising that the intrinsic width of the resonances increases more rapidly than the thermal broadening of the electron distribution function in a temperature range for which phase coherence between the resonant and non-resonant channels is preserved.  We plan extensive measurements, varying temperature and magnetic field as well as drain-source voltage in the weakly-coupled regime, to shed light on the nature of the continuous transmission that gives rise to the Fano interference and on the origin of the thermal broadening of the resonances.

One of us (G.~G.) acknowledges support from the National Sciences and Engineering Research Council of Canada. This work was supported by the US Army Research Office under contract DAAG~55-98-1-0138 and by the National Science Foundation under grant number DMR-9732579.


\vspace{-.25in}

\begin{thebibliography}{10}

\bibitem[\ddagger]{davidaddress}
Current address: Harvard University, Department of Physics and Society of
Fellows, 17 Oxford Street, Cambridge MA 02138.

\bibitem[*]{byline}
{\it mkastner@mit.edu} \newline

\bibitem{kastner}
M. A. Kastner, Physics Today {\bf 46}, 24 (1993).

\bibitem{ashoori}
R. C. Ashoori, Nature {\bf 379}, 413 (1996).

\bibitem{Gores} J. G\protect{\"{o}}res, D. Goldhaber-Gordon, S. Heemeyer, M. A. Kastner, H. Shtrikman, D. Mahalu and U. Meirav, Phys. Rev. B {\bf62}, 2188 (2000).

\bibitem{david}
D. Goldhaber-Gordon, H. Shtrikman, D. Mahalu, D. Abusch-Magder, U.
Meirav, and M. A. Kastner, Nature {\bf 391}, 156 (1998).

\bibitem{davidprl}
D. Goldhaber-Gordon, J. G\"{o}res, M. A. Kastner, H. Shtrikman, D.
Mahalu, and U. Meirav, Phys. Rev. Lett. {\bf 81}, 5225 (1998).

\bibitem{fano}
U. Fano, Phys. Rev. {\bf 124}, 1866 (1961).


\bibitem{meirav}
U. Meirav, M. A. Kastner, and S. J. Wind, Phys. Rev. Lett. {\bf 65}, 771 (1990).

\bibitem{foxmancross}
E. B. Foxman, U. Meirav, P. L. McEuen, M. A. Kastner, O. Klein, P. A. Belk, D. M. Abusch and S. J. Wind, Phys. Rev. B {\bf 50}, 14193 (1994).

\bibitem{foxman}
E. B. Foxman, P. L. McEuen, U. Meirav, N. S. Wingreen, Y. Meir, P. A.
Belk, N. R. Belk, M. A. Kastner, and S. J. Wind, Phys. Rev. B {\bf 47}, 10020 (1993).

\bibitem{averin}
D. V. Averin and A. A. Odintsov,  Physics Letters A {\bf 140}, 251 (1989).

\bibitem{beenakker}
C. W. J. Beenakker, Phys. Rev. B {\bf44}, 1646 (1991).

\bibitem{stone}
A. D. Stone and P. A. Lee, Phys. Rev. Lett. {\bf54}, 1196 (1985).

\bibitem{buttiker}
M. B\protect{\"{u}}ttiker, Phys. Rev. {\bf33}, 3020 (1986).

\bibitem{sivan}
U. Sivan, Y. Imry and A. G. Aronov, Europhys. Lett. {\bf28}, 115 (1994).

\bibitem{huibers}
A. G. Huibers, M. Switkes, C. M. Marcus, K. Campman and A. C. Gossard, Phys. Rev. Lett. {\bf81}, 200 (1998).

\bibitem{averin}
D. V. Averin and Yu. V. Nazarov, Phys. Rev. Lett. {\bf65}, 2446 (1990).

\bibitem{li}
Jiutao Li, Wolf-Dieter Schneider, Richard Berndt, and Bernard Delley, Phys. 
Rev. Lett. {\bf80}, 2893 (1998).

\bibitem{madhavan}
V. Madhavan, W. Chen, T. Jamneala, M. F. Crommie, and N. S. Wingreen, Science
{\bf280}, 567 (1998).

\end{thebibliography}
\end{document}